\documentclass[conference]{IEEEtran}
\IEEEoverridecommandlockouts

\usepackage{cite}
\usepackage{amsmath,amssymb,amsfonts}
\usepackage{graphicx}
\usepackage{textcomp}
\usepackage{xcolor}
\usepackage{booktabs}
\usepackage{multirow}
\usepackage{enumitem}
\usepackage{url}

\setlist[enumerate]{leftmargin=*, itemsep=2pt, topsep=2pt}

\usepackage{stfloats} 
\begin{document}
\bstctlcite{IEEEexample:BSTcontrol}

\title{Beyond Text: Text--Image Fusion for Scalable Two-Tower Retrieval}

\author{
\IEEEauthorblockN{Qujiaheng Zhang\textsuperscript{*}}
\IEEEauthorblockA{Target \\
Minneapolis, MN, USA \\
zhan7977@alumni.umn.edu}
\and
\IEEEauthorblockN{Guangyue Xu\textsuperscript{*}}
\IEEEauthorblockA{Target \\
Minneapolis, MN, USA \\
guangyue.xu@target.com}
\and
\IEEEauthorblockN{Fengjie Li}
\IEEEauthorblockA{Target \\
Sunnyvale, CA, USA \\
alex.li@target.com}
\thanks{\textsuperscript{*}Both authors contributed equally to this work.}
}

\maketitle

\begin{abstract}
E-commerce customers make purchase decisions by jointly considering product text and visual information, yet most industrial retrieval systems rely almost exclusively on textual signals. Deployed at scale on CPU-based nearest-neighbor infrastructure, these systems leave substantial visual evidence untapped. In this work, we study unified text--image fusion for two-tower retrieval models at a large-scale e-commerce platform, training on 20M query--item pairs.
We demonstrate that domain-specific fine-tuning and modality-specific query alignment between user queries and both product text and image modalities are crucial for effective multimodal retrieval.
Building on these insights, we propose a mixture-of-modality-experts fusion architecture with a bilinear interaction network to capture cross-modal complementary information.
Our model achieves up to 4.86\% improvement in nDCG@1 for desirability and 2.36\% for relevance over a strong text-only baseline, while preserving the pre-computable
item embedding design required for production deployment.
\end{abstract}


\section{Introduction}

\begin{figure}[t]
    \centering
    \includegraphics[width=0.95\columnwidth]{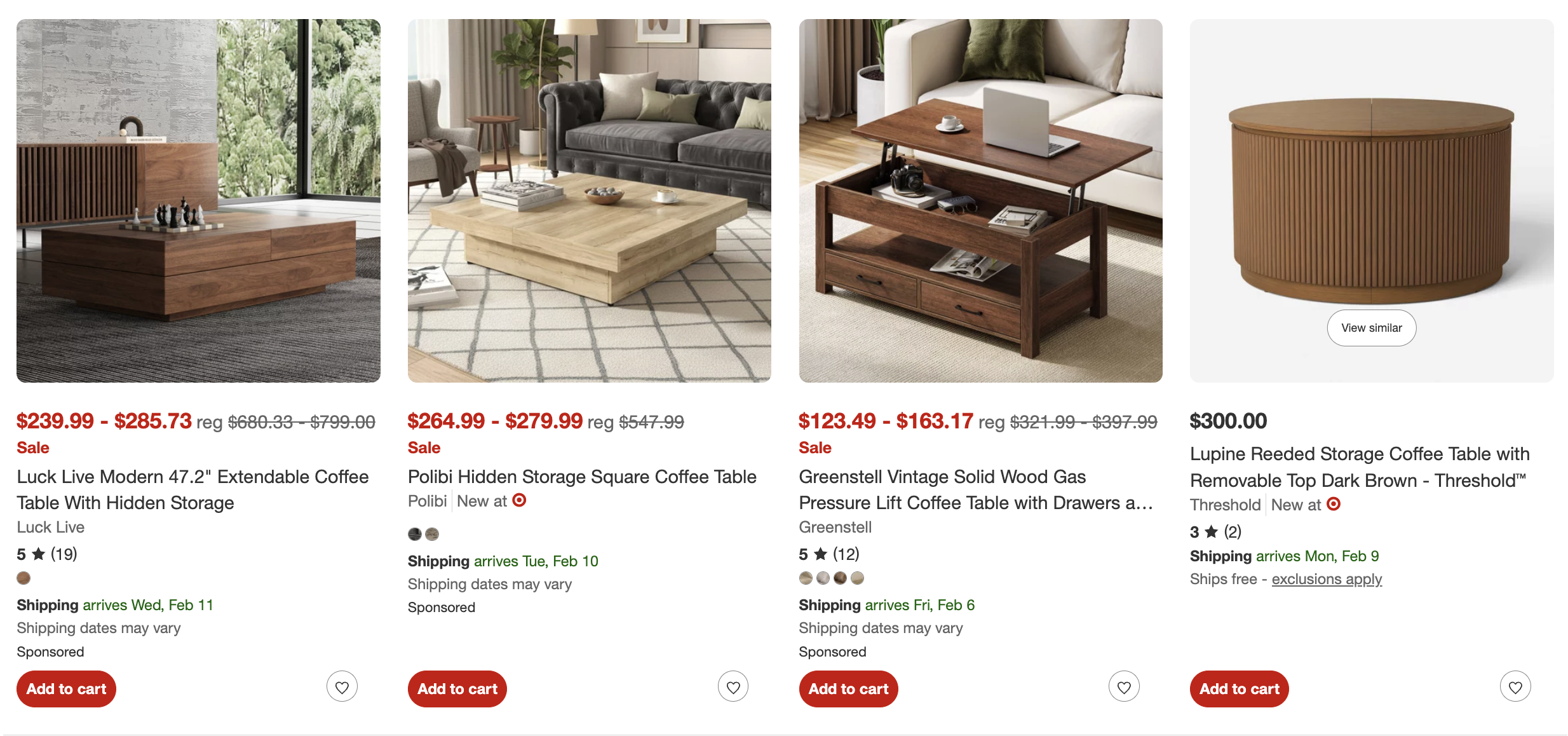}
    \caption{Home decor search is highly visual: product images provide critical relevance
    signals beyond text, motivating multimodal retrieval.}
    \label{fig:search_results}
\end{figure}

Large-scale e-commerce retrieval systems are typically optimized around textual relevance,
encoding user queries and product descriptions into a shared embedding space to enable efficient nearest-neighbor search at scale.
While this design is computationally attractive and widely adopted in industry~\cite{walmart_semantic_retrieval_22_kdd, target_unified_ranking_26}, it fails to
reflect the inherently multimodal nature of user decision-making in online shopping. In
practice, as illustrated in Fig.~\ref{fig:search_results}, users often rely on visual
information---such as appearance, style, color, and fine-grained details conveyed by product images---to assess relevance and make interaction decisions, especially when textual descriptions are sparse, ambiguous, or insufficient. This mismatch between the multimodal evidence users consider and the unimodal signals retrieval models are trained on leads to suboptimal relevance modeling, causing text-only retrieval methods to struggle in visually driven categories where subtle visual distinctions dominate relevance judgments. As product catalogs grow and titles are increasingly written for search-engine coverage rather than natural description, the gap between what a title says and what a shopper actually sees only widens, making this a compounding rather than a static problem. At the scale this system operates---on the order of millions of indexed products and production query volumes reaching thousands of queries per second (Section~\ref{sec:deployment})---even a small per-query gap between text and true product relevance compounds into a substantial loss of retrieval quality, motivating the incorporation of visual signals that prior work has begun to explore.
Prior work has shown that incorporating visual signals benefits product retrieval~\cite{PictureWorthWords,MultimodalSemanticRetrieval} and that multimodal pre-training improves search quality at scale~\cite{bringing_multimodality_amazon}.

Despite substantial progress in multimodal representation learning, applying multimodal
techniques to e-commerce retrieval poses domain-specific challenges. VL-CLIP~\cite{Vl-clip}
points out weak object-level alignment and ambiguous textual representations in product data;
EI-CLIP~\cite{Ei-clip} identifies domain-specific entities as confounders and introduces causal
interventions; FashionKLIP~\cite{Fashionklip} leverages external knowledge graphs to address the
lack of fine-grained cross-modal conceptual knowledge. While these approaches improve
multimodal understanding, many introduce additional architectural complexity or computational
overhead that makes large-scale nearest-neighbor retrieval
challenging~\cite{UniECS, Category2Image}. This motivates us to build an efficient multimodal
retriever that maintains scalable two-tower search, supports large-scale indexing, and can be
deployed in CPU-based production infrastructure. Concretely, we require an architecture in which
adding the image modality does not change the online serving path at all: item embeddings must
remain precomputable offline, the query tower must remain a single lightweight text encoder, and
nearest-neighbor retrieval must remain a drop-in operation over the same ANN index used by the
text-only system today.

In this work, we directly address the mismatch between user decision signals and model learning
signals in e-commerce retrieval. Our experiments demonstrate consistent improvements of up to
4.86\% in nDCG@1 over a strong text-only baseline, confirming that visual signals provide
complementary evidence that text alone cannot capture. To achieve these gains, we show that
domain-specific fine-tuning is essential for aligning multimodal representations with commercial
relevance, and that explicitly aligning user queries with both product text and product images
enables the model to learn from the same visual cues that users rely on during relevance
judgment. We further propose a lightweight mixture-of-modality-experts fusion architecture
combined with bilinear interaction to model fine-grained cross-modal relationships, and
introduce a multi-stage curriculum training strategy with a multi-objective loss that jointly
models user engagement and semantic relevance.

This work makes the following contributions to large-scale e-commerce retrieval, trained and evaluated at Big Data scale on 20M query--item pairs and an 8-million-item production catalog: 
\begin{itemize}
    \item We demonstrate that visual signals provide significant and consistent retrieval
    improvements, especially for visually driven product categories, and identify domain
    fine-tuning and modality-specific query alignment as the key enablers.
    \item We propose a mixture-of-modality-experts fusion architecture with a bilinear
    interaction network that enables effective integration of textual and visual representations
    for scalable two-tower retrieval.
    \item We introduce a multi-stage curriculum training strategy with a multi-objective loss
    that jointly models user engagement and semantic relevance, and demonstrate consistent
    improvements across both engagement-based and relevance-based offline evaluation benchmarks.
\end{itemize}

\begin{figure*}[!t]
    \centering
    \includegraphics[width=0.9\textwidth]{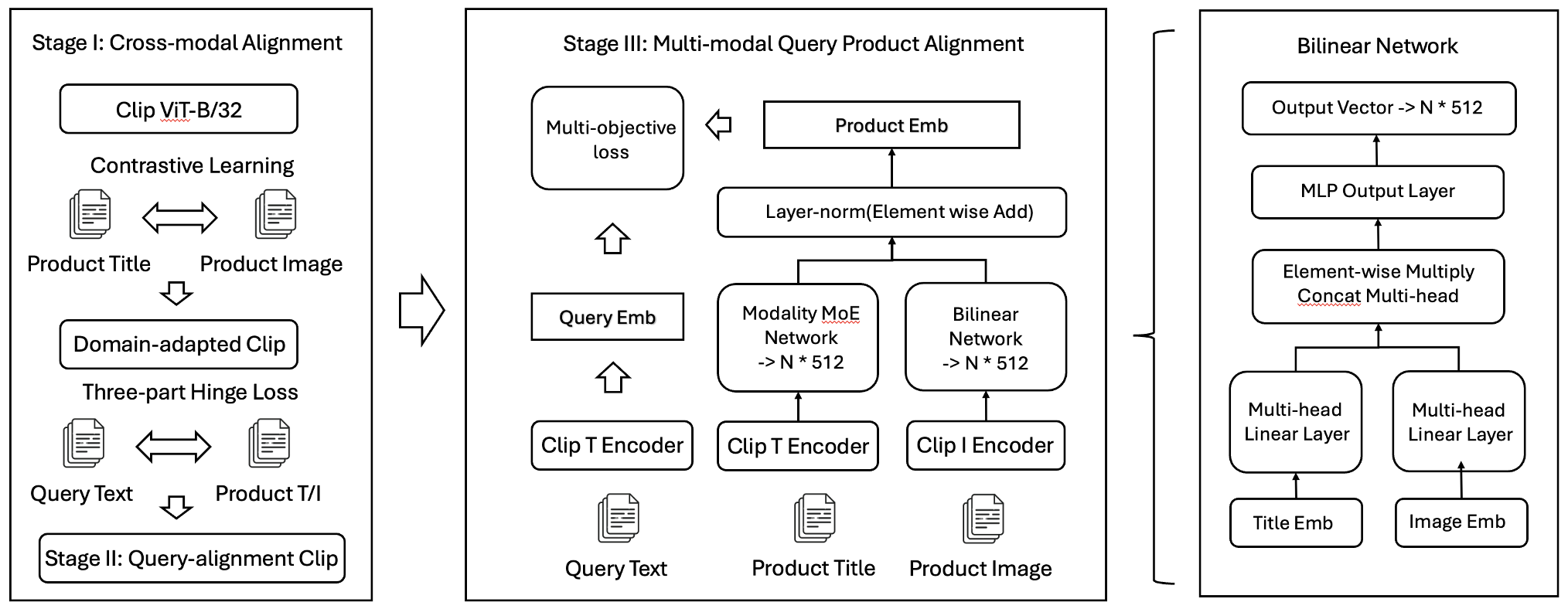}
    \caption{Proposed multimodal retrieval framework.
    Left: curriculum training procedure for domain adaptation and query alignment;
    Middle: multimodal two-tower retrieval architecture;
    Right: bilinear block architecture.}
    \label{fig:architecture}
\end{figure*}

\section{Related Work}

\noindent\textbf{Multimodal E-Commerce Retrieval.}
Industrial retrieval systems have long relied on text-only two-tower models, but growing
evidence shows that product images carry complementary relevance signals that text alone cannot
capture~\cite{PictureWorthWords,MultimodalSemanticRetrieval}. This is particularly acute in
e-commerce, where product titles are often optimized for keyword coverage rather than for
faithfully describing appearance, and where the same textual attributes (e.g., ``blue,''
``modern'') can correspond to visually disparate products depending on category~\cite{xu2021zscl,xu2023metarevision,xu2024thesis}. Early work on adapting
CLIP~\cite{radford2021clip,xu2022prompting,xu2024gipcol} to e-commerce domains includes VL-CLIP~\cite{Vl-clip}, which
addresses weak object-level alignment via LLM-augmented embeddings; EI-CLIP~\cite{Ei-clip},
which introduces causal interventions for entity-level disambiguation; and
FashionKLIP~\cite{Fashionklip}, which leverages knowledge graphs for fine-grained cross-modal
alignment. CommerceMM~\cite{Yu_KDD_2022_CommerceMM} and MSRA~\cite{Jin_CIKM_2023_MSRA}
demonstrate the value of large-scale multimodal pre-training for product search. Amazon's visual
search system~\cite{bringing_multimodality_amazon} shows that multimodal signals improve
retrieval in production at scale. Our work differs by focusing on a lightweight, deployable
two-tower architecture with systematic ablation of fusion strategies under a curriculum training
regime.

\noindent\textbf{Two-Tower Retrieval.}
Two-tower models~\cite{walmart_semantic_retrieval_22_kdd, target_unified_ranking_26} are the
dominant paradigm for large-scale retrieval due to their efficiency: item embeddings are
pre-computed offline and queries are encoded at serving time, enabling sub-100\,ms latency at
billions of items. Extending this paradigm to multimodal inputs requires careful design to
maintain scalability. Fusing modalities at the item tower---rather than at query time---preserves
the pre-computation property and keeps serving latency identical to the text-only baseline,
which motivates our fusion-at-item-tower approach. This design also allows the heavier image
encoder to run entirely offline, making the system practical for CPU-based production
infrastructure. In contrast, fusing modalities at query time would require running the image
encoder---or a comparably expensive cross-modal scorer---on every incoming query, which is
difficult to reconcile with the latency and cost budgets of CPU-based serving at billions of
items; it also forecloses precomputing and indexing item embeddings ahead of time, which is the
property that makes two-tower retrieval tractable at this scale in the first place.

\noindent\textbf{Curriculum and Multi-Objective Training.}
Curriculum learning for dense retrieval~\cite{he2023capstone, lin2023prod} and hard negative
mining~\cite{zhou2022simans} have shown consistent gains in text retrieval by progressively
increasing training difficulty. We adapt these ideas to the multimodal setting: our three-stage
curriculum first adapts encoders to the e-commerce domain, then aligns each modality with user
queries independently, and finally trains the full fusion model---a progression that stabilizes
optimization and improves final retrieval quality. Taken together, these three lines of work point to a common gap: prior multimodal e-commerce systems either add architectural complexity that undermines two-tower scalability, or, where two-tower efficiency is preserved, do not systematically study how domain adaptation, query alignment, and fusion design interact. The method we present next is designed to close that gap directly.

\section{Proposed Method}

We now describe our approach in four parts. Section~\ref{sec:task_formulation} formalizes the
retrieval objective; Section~\ref{sec:model_architecture} describes the two-tower architecture and
the mixture-of-modality-experts fusion module with bilinear cross-modal interaction;
Section~\ref{sec:neg_sampling} describes the negative sampling strategy used during training; and
Sections~\ref{sec:training_objective}--\ref{sec:curriculum_training} describe the multi-objective
loss and the curriculum training strategy that together turn this architecture into an effective
multimodal retriever.

\subsection{Task Formulation}
\label{sec:task_formulation}

Given a user query $q$ and a large candidate set of products
$\mathcal{X}=\{x_1,\dots,x_n\}$, the goal of e-commerce retrieval is to return the items that
best match the query. Depending on the application, matching may reflect semantic relevance
between the query and the product, implicit user engagement signals such as clicks and
purchases, or a combination of the two. In our setting, each product $x \in \mathcal{X}$ is
represented by a textual description such as its title $t$ and a corresponding product image
$i$, which provide complementary semantic and visual evidence for matching estimation.

We aim to learn a function $f(q,x)$ that measures the overall matching degree between a query
and a candidate product, and induces a ranking over the candidate set $\mathcal{X}$:
\begin{equation}
x^{*} = \arg\max_{x \in \mathcal{X}} \; f(q,x).
\label{eq:retrieval_objective}
\end{equation}
The optimal item $x^*$ is retrieved via efficient nearest-neighbor search in the shared
embedding space. The remainder of this section describes how we instantiate $f(q,x)$ with a
two-tower architecture whose item tower fuses text and image evidence (Section~\ref{sec:model_architecture}), how negatives are sampled during training (Section~\ref{sec:neg_sampling}), and the multi-objective curriculum used to train it (Sections~\ref{sec:training_objective}--\ref{sec:curriculum_training}).

\subsection{Model Architecture}
\label{sec:model_architecture}

We adopt a standard two-tower retrieval framework for scalable candidate generation in
large-scale e-commerce search. The query tower encodes a user query $q$ into an embedding
$\mathbf{h}_q$, while the item tower maps each product $x$ into the same semantic space for
efficient nearest-neighbor retrieval with cosine similarity.

\textbf{Query and Item Encoding.}
We initialize both text and image representations from a pretrained CLIP
model~\cite{radford2021clip}. We encode query $q$ via the CLIP text encoder as
$\mathbf{h}_q=f_{\text{text}}(q)$, and encode each product's title $t$ and image $i$ as
$\mathbf{h}_t=f_{\text{text}}(t)$ and $\mathbf{h}_i=f_{\text{img}}(i)$.

\textbf{Mixture-of-Modality Fusion.}
We integrate complementary semantic and visual signals via a lightweight
mixture-of-modality-experts fusion module. A gating network---a two-layer MLP applied to the
concatenation of $\mathbf{h}_t$ and $\mathbf{h}_i$---predicts an adaptive weight
$\alpha\in[0,1]$, and we compute the fused item representation as:
\begin{equation}
\mathbf{h}_f = \alpha \mathbf{h}_t + (1-\alpha)\mathbf{h}_i .
\end{equation}

\textbf{Cross-modal Interaction.}
To capture fine-grained feature interactions beyond linear fusion, we introduce a multi-head
bilinear interaction network~\cite{fukui2016mcb,benyounes2017mutan}. Given the text embedding
$\mathbf{h}_t$ and image embedding $\mathbf{h}_i$, we apply $K$ learnable projection heads that
combine features via element-wise multiplication, concatenate their outputs across all heads, and
pass them through a lightweight MLP. The interaction output is combined with $\mathbf{h}_f$ via
a residual connection and layer normalization to yield the final item embedding:
\begin{equation}
\mathbf{h}_x =
\text{LayerNorm} \!\left(
\mathbf{h}_f +
\text{MLP}\!\Big(
\big\Vert_{k=1}^{K}
\bigl(
\mathbf{W}_t^{(k)} \mathbf{h}_t
\odot
\mathbf{W}_i^{(k)} \mathbf{h}_i
\bigr)
\Big)
\right).
\label{eq:item_emb}
\end{equation}
Query--item relevance is then measured by cosine similarity:
\begin{equation}
s(q,x)=\cos(\mathbf{h}_q,\mathbf{h}_x).
\end{equation}

\subsection{Self-Adversarial Negative Sampling}
\label{sec:neg_sampling}

We adopt a self-adversarial negative sampling strategy within each mini-batch. Given a batch of
$N$ query--product pairs, we compute the cosine similarity matrix between query and product
embeddings, where diagonal entries correspond to positives and off-diagonal entries provide
candidate negatives. Writing $s_{jk}$ for the similarity between query $j$ and product $k$, we
draw $N$ negatives per batch (one per query on average) without replacement from the
off-diagonal entries with probability
\begin{equation}
p_{jk} = \frac{\max(0, s_{jk})}{\sum_{j' \neq k'} \max(0, s_{j'k'})}, \qquad j \neq k,
\label{eq:neg_sampling}
\end{equation}
which avoids the instability of always selecting the hardest negatives while still
concentrating sampling probability on more confusable items. As encoders
improve during training, the similarity distribution naturally sharpens around harder negatives,
providing an automatic curriculum over difficulty~\cite{instcartebr,zhou2022simans,he2023capstone}
without any hyperparameter changes.

\subsection{Training Objective}
\label{sec:training_objective}

We train the model with two complementary supervision signals: a desirability (engagement) label
$y_{\text{eng}}$ derived from aggregated user interactions and a semantic relevance label
$y_{\text{rel}}$ derived from human-aligned cross-encoder scores. Both are graded categorical
signals with three levels: high interaction ($y=2$), low interaction ($y=1$), and no interaction
($y=0$).

We adopt a \emph{multi-objective} formulation: unlike a multi-task approach that routes each
label to a separate output head, we apply both supervision signals to the \emph{same} predicted
score $\hat{y} = s(q,x)$. This design keeps the two-tower architecture intact, preserving
pre-computable item embeddings and standard nearest-neighbor retrieval---while allowing the
unified embedding space to absorb information from heterogeneous label sources.

Given $\hat{y}$ and a label $y$, we adopt a three-part squared-hinge loss:
\begin{equation}
\begin{aligned}
\mathcal{L}_{3-\text{hinge}}(\hat{y}, y)
= \;& I^{+}(y)\,\max(0, \epsilon_{+} - \hat{y})^{m} \\
&+ I^{-}(y)\,\max(0, \hat{y} - \epsilon_{-})^{m} \\
&+ I^{0}(y)\,\max(0, \hat{y} - \epsilon_{0})^{m},
\end{aligned}
\label{eq:3hinge}
\end{equation}
where $I^{+}(y) = \mathbf{1}[y\!=\!2]$, $I^{-}(y) = \mathbf{1}[y\!=\!1]$, and
$I^{0}(y) = \mathbf{1}[y\!=\!0]$. The ordered margins $\epsilon_{+} > \epsilon_{-} > \epsilon_{0}$
encode the desired ranking, leaving an explicit dead zone between adjacent levels so that graded
labels are not collapsed into a binary contrast. The exponent $m$ controls how sharply margin
violations are penalized. We set $\epsilon_{+}=0.8$, $\epsilon_{-}=0.65$, $\epsilon_{0}=0.4$,
and $m=2$, so that large violations are penalized quadratically while small ones contribute mildly.

We combine both label sources in the final objective:
\begin{equation}
\mathcal{L}_{\text{total}}
=\lambda_{\text{eng}}\,\mathbb{E}\!\left[\mathcal{L}_{3-\text{hinge}}(\hat{y}, y_{\text{eng}})\right]
+\lambda_{\text{rel}}\,\mathbb{E}\!\left[\mathcal{L}_{3-\text{hinge}}(\hat{y}, y_{\text{rel}})\right],
\label{eq:total}
\end{equation}
with $\lambda_{\text{eng}}=0.7$ and $\lambda_{\text{rel}}=0.3$, prioritizing engagement as the
primary business objective while using relevance as a regularizer that anchors the embedding
space to semantic correctness when engagement signals are sparse or noisy. We ablate this
weighting choice empirically in Section~\ref{sec:ablations} (Table~\ref{tab:weight_ablation}).

\subsection{Curriculum Training Strategy}
\label{sec:curriculum_training}

We adopt a curriculum-style multi-stage training strategy to progressively align model
representations with user decision signals in e-commerce
search~\cite{he2023capstone,lin2023prod}. We decompose the learning process into three stages
with increasing supervision complexity, allowing the model to gradually acquire modality-aware
and cross-modal relevance patterns.

\textbf{Stage~I: Domain Adaptation of CLIP Encoders.}
We fine-tune the CLIP encoders on a large-scale title--image dataset constructed from Target
product information, using contrastive learning between product titles and product images to
adapt pretrained visual-text representations to e-commerce semantics. We adopt the
compositional optimization framework of FastCLIP~\cite{wei2024fastclip}, which maintains a
moving-average negative sample queue (memory bank) to approximate the full-corpus negative
distribution and enables efficient global contrastive loss optimization without requiring an
extremely large batch size.

\textbf{Stage~II: Modality-Specific Query Alignment.}
We explicitly align user queries with individual item modalities \emph{sequentially}---one
epoch per modality---using the three-part hinge loss. This step encourages the model to learn
modality-specific relevance signals while incorporating graded ranking supervision, reducing the
mismatch between user intent and item representations.

\textbf{Stage~III: Multimodal Fusion Alignment.}
We align queries with the fused product representations produced by the mixture-of-modality-experts
architecture, enabling the model to learn unified multimodal embeddings that capture both
unimodal evidence and cross-modal interactions, resulting in more effective retrieval and
improved downstream ranking performance.

\section{Experiments}

We conduct experiments to validate the three key design choices in our method: domain
adaptation (Stage~I), query alignment (Stage~II), and fusion architecture (Stage~III). We also
evaluate the overall impact of incorporating visual signals and analyze the learned fusion
behaviors. Concretely, we structure our evaluation around four questions: (1) does incorporating
product images improve retrieval quality over a strong text-only baseline, and is the improvement
statistically robust (Section~\ref{sec:results})? (2) where do we see adding image helps and on which queries does visual signal help
most (Section~\ref{sec:visual_signal_analysis})? (3) how much
of the gain is attributable to domain fine-tuning versus query alignment versus fusion
architecture (Section~\ref{sec:ablations})? and (4) can the resulting model be served within the
latency and cost constraints of production CPU-based infrastructure (Section~\ref{sec:deployment})?

\subsection{Training Dataset}

We construct our training data from three months of e-commerce search logs, yielding the top
20M aggregated query--item pairs. Each item is represented by its product title and image.
We derive a desirability label from aggregated interaction signals (clicks, add-to-cart,
purchases) and an auxiliary semantic relevance label from a cross-encoder trained on
human-annotated data. The desirability score is computed as
$1\cdot\text{CTR} + 5\cdot\text{ATC rate} + 10\cdot\text{purchase rate}$ and bucketed into
low ($y_{\text{eng}}=1$; mostly clicks with few add-to-carts) and high ($y_{\text{eng}}=2$;
all three interaction types present). Negatives ($y_{\text{eng}}=0$) are additionally sampled
items labeled as not desirable. The label distribution is approximately 1:3:6 for high, low, and
no-interaction labels, reflecting the natural sparsity of strong engagement signals in
e-commerce search logs.

\subsection{Evaluation Dataset}

We evaluate on two held-out benchmarks capturing complementary aspects of retrieval quality.
To prevent leakage, all evaluation queries are removed from the training set.

\textbf{Desirability.}~
A log-based dataset from a disjoint time window reflecting graded user engagement. It contains
247{,}895 query--item pairs over 6{,}284 queries, with 39 items per query on average.

\textbf{Relevance.}~
A human-annotated dataset providing gold-standard semantic judgments. It includes 393{,}425
pairs from 4{,}975 queries, with an average of 79 items per query. Queries whose candidates are
all labeled 0 are excluded, since nDCG is undefined when the ideal DCG is zero.

Both evaluation sets use four grades, one level finer than the three training levels in
Section~\ref{sec:training_objective}: desirability is graded 1--4 and relevance 0--3. The model
never predicts these grades; it only produces a score $s(q,x)$ used for ranking, and nDCG uses
the graded labels directly as gains.

\subsection{Experimental Setup}

We evaluate a series of two-tower retrieval variants to quantify the contribution of visual
signals, alignment strategies, and fusion architectures. All multimodal models are built upon
CLIP-based text and image encoders, differing only in their adaptation and fusion designs:

\begin{itemize}[leftmargin=1.2em, itemsep=0pt, topsep=0pt, parsep=0pt]
    \item \textbf{Text-only (Baseline).} CLIP text encoder over queries and product titles,
    fine-tuned with the multi-objective loss. This is the text-only baseline used in
    Table~\ref{tab:image_impact}.
    \item \textbf{Pretrained CLIP (MoE Fusion).} Multimodal retriever with pretrained CLIP
    encoders and MoE fusion, jointly fine-tuned during retrieval training. This serves as the
    starting point for alignment ablations in Table~\ref{tab:alignment_ablation}.
    \item \textbf{+FT.} CLIP encoders additionally adapted via Stage~I domain fine-tuning on 4M
    title--image pairs.
    \item \textbf{+FT+Align.} Further adds Stage~II modality-specific query alignment.
    \item \textbf{Fusion Ablations.} All fusion variants (MLP, MoE, MoE+MLP, Attention,
    MoE+Bilinear) share the same +FT+Align encoder initialization and differ only in their
    fusion module design.
\end{itemize}

\textbf{Training Infrastructure.}
All stages run on two NVIDIA A100 GPUs (batch size 512, AdamW optimizer, weight decay
$10^{-4}$, linear learning-rate schedule with 200 warm-up steps). Encoders are initialized from
CLIP ViT-B/32, and the query and title share one text encoder. Stage~I runs for
5 epochs and completes in approximately 4 hours. Stage~II runs for 1 epoch per modality at
learning rate $10^{-5}$, totaling about 3 hours across the two modalities. Stage~III runs for
6 epochs at learning rate $2\times10^{-5}$ and takes roughly 11 hours. The bilinear module uses
$K=4$ heads of 128 dimensions each, and the gating network maps the 1024-d concatenation to a
256-d hidden layer. The full curriculum thus trains end-to-end in under 18 hours of wall-clock
time on a 2$\times$A100 setup, making the framework reproducible within a single-day training
budget.

\subsection{Metrics}

We evaluate retrieval performance using normalized Discounted Cumulative Gain (nDCG@K):
\begin{equation}
\mathrm{DCG@K} = \sum_{i=1}^{K} \frac{2^{r_i}-1}{\log_2(i+1)}, \qquad
\mathrm{nDCG@K} = \frac{\mathrm{DCG@K}}{\mathrm{IDCG@K}},
\end{equation}
where $r_i$ denotes the label of the item at rank $i$ and $\mathrm{IDCG@K}$ is the DCG of the
ideal ranking. We adopt nDCG rather than a purely position-agnostic metric because the retrieval
model's output is consumed directly as the final ranking in production, with no downstream
re-ranking stage to further reorder candidates; nDCG's rank-discounted formulation therefore
reflects the metric that matters at serving time.

\subsection{Experimental Results}
\label{sec:results}

We first evaluate the overall impact of incorporating product image information into two-tower retrieval models. Table~\ref{tab:image_impact} compares the text-only baseline with the proposed MoE+Bilinear
model. The multimodal model yields consistent improvements across all nDCG cutoffs for both
desirability and relevance. The gains are largest at the top of the ranking---4.86\% at nDCG@1
for desirability and 2.36\% at nDCG@1 for relevance---and diminish at deeper cutoffs. This
pattern is consistent with the hypothesis that visual signals are most discriminative for
distinguishing the single best match from near-miss candidates, where textual descriptions alone
are insufficient to resolve fine-grained visual differences. We next assess the statistical
significance of this gain and examine which queries benefit most, before ablating the individual
design choices behind it.

\textbf{Statistical Significance.}
To verify these gains are systematic rather than an artifact of a few queries, we conduct a
paired Wilcoxon signed-rank test on the per-query $\Delta$nDCG@24 between the MoE+Bilinear model
and the text-only baseline, together with 95\% confidence intervals from 10{,}000 bootstrap
resamples. On Desirability ($n=6{,}284$), nDCG@24 improves by 1.22pp with
p value $<$ 0.001, with the multimodal model winning on 57.7\% of queries versus 36.3\% for the baseline
(6.0\% ties). On Relevance ($n=4{,}975$), nDCG@24 improves by 0.86pp with 
p value $<$0.001, with a 42.2\% vs.\ 31.9\% win rate (25.9\% ties). Both gains are highly significant,
with confidence intervals well clear of zero, indicating the improvement is consistent across the
query population rather than concentrated in a small subset of queries.

\begin{table}[t]
\centering
\footnotesize
\caption{Effect of incorporating product images on retrieval performance (nDCG).}
\label{tab:image_impact}
\setlength{\tabcolsep}{3pt}
\begin{tabular}{lcccccccc}
\toprule
\multirow{2}{*}{\textbf{Method}}
& \multicolumn{4}{c}{\textbf{Desirability}}
& \multicolumn{4}{c}{\textbf{Relevance}} \\
\cmidrule(lr){2-5} \cmidrule(lr){6-9}
& @1 & @3 & @9 & @24 & @1 & @3 & @9 & @24 \\
\midrule
Baseline
& 0.802 & 0.809 & 0.839 & 0.895
& 0.890 & 0.888 & 0.890 & 0.897 \\
MoE+Bilinear
& \textbf{0.841} & \textbf{0.839} & \textbf{0.861} & \textbf{0.909}
& \textbf{0.911} & \textbf{0.905} & \textbf{0.903} & \textbf{0.909} \\
\midrule
$\Delta$\%
& +4.86 & +3.71 & +2.62 & +1.56
& +2.36 & +1.91 & +1.46 & +1.34 \\
\bottomrule
\end{tabular}
\end{table}

\subsection{When Does Visual Signal Help Most?}
\label{sec:visual_signal_analysis}

To understand \emph{where} visual signal is most informative, we conduct a diagnostic analysis
comparing the full multimodal model against the text-only model---the same two systems compared
in Table~\ref{tab:image_impact}---broken down by query intent. We classify each query into one of five intents by simple lexical rules:
\emph{branded} (query contains the target brand), \emph{visual\_attribute} (query names a color,
material, or finish~\cite{xu2021zscl,xu2024gipcol}), \emph{functional\_benefit} (query names a use-case or property),
\emph{product\_type} (query overlaps with the target item type), and \emph{generic\_broad}
(fallback).

Table~\ref{tab:query_intent} reports the mean per-query $\Delta$nDCG@10 (multimodal $-$ text-only)
and multimodal win rate by intent on Desirability. \emph{visual\_attribute} queries benefit the
most (+0.0230, 56.7\% win rate), while \emph{branded} queries benefit the least among
well-populated buckets (+0.0130, 54.1\%)---consistent with brand names being resolved through
literal text match rather than visual appearance. The visual-signal advantage also grows with the
difficulty of the ranking task: mean $\Delta$nDCG@10 rises from +0.010 for query pools spanning
only two distinct label values to +0.028 for pools spanning all four levels, and from +0.010 for
pools with no other relevant items to +0.029 for pools with ten or more---image similarity is most
valuable exactly when many visually and textually similar candidates must be finely
discriminated.

\begin{table}[t]
\centering
\footnotesize
\caption{Comparison of the multimodal model vs.\ the text-only model by query intent, on
Desirability.}
\label{tab:query_intent}
\setlength{\tabcolsep}{3pt}
\begin{tabular}{lccc}
\toprule
\textbf{Query Intent} & \textbf{n} & \textbf{$\Delta$nDCG@10} & \textbf{Multimodal Win Rate} \\
\midrule
visual\_attribute            & 261   & \textbf{+0.0230} & \textbf{56.7\%} \\
generic\_broad                & 2{,}407 & +0.0188          & 56.5\% \\
product\_type                 & 2{,}293 & +0.0189          & 55.9\% \\
branded                       & 1{,}224 & +0.0130          & 54.1\% \\
functional\_benefit$^\dagger$ & 99    & +0.0132          & 50.5\% \\
\bottomrule
\end{tabular}
\vspace{2pt}

\raggedright\footnotesize $^\dagger$Small sample; interpret with caution.
\end{table}

\begin{figure}[t]
\centering
\includegraphics[width=0.95\columnwidth]{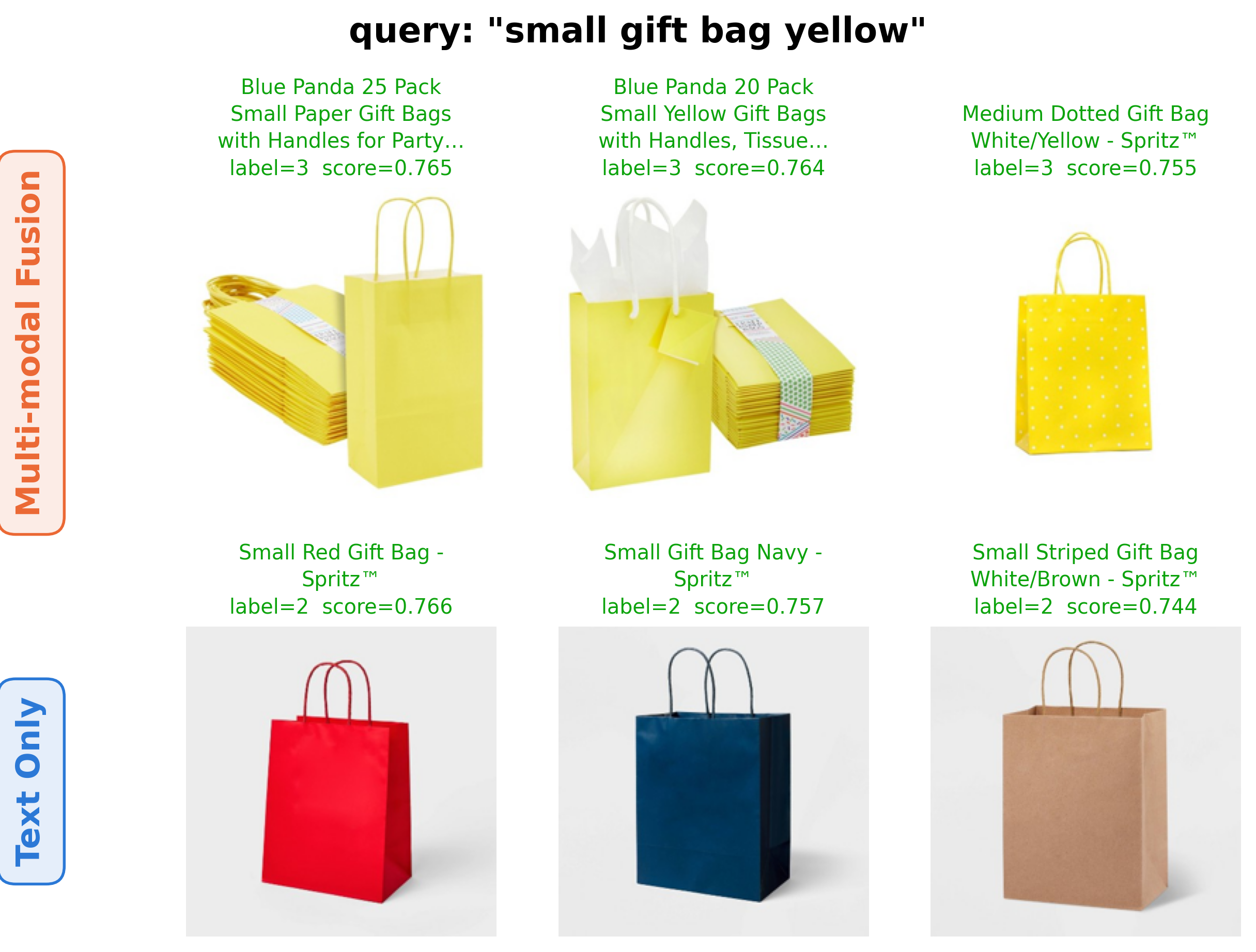}
\caption{Query \emph{``small gift bag yellow''} (visual\_attribute). The multimodal model returns
yellow gift bags (label~3); the text-only model matches the correct product type but the wrong
color (red/navy, label~2)---a case where color, not text, resolves relevance.}
\label{fig:qual_giftbag}
\end{figure}

\begin{figure}[t]
\centering
\includegraphics[width=0.95\columnwidth]{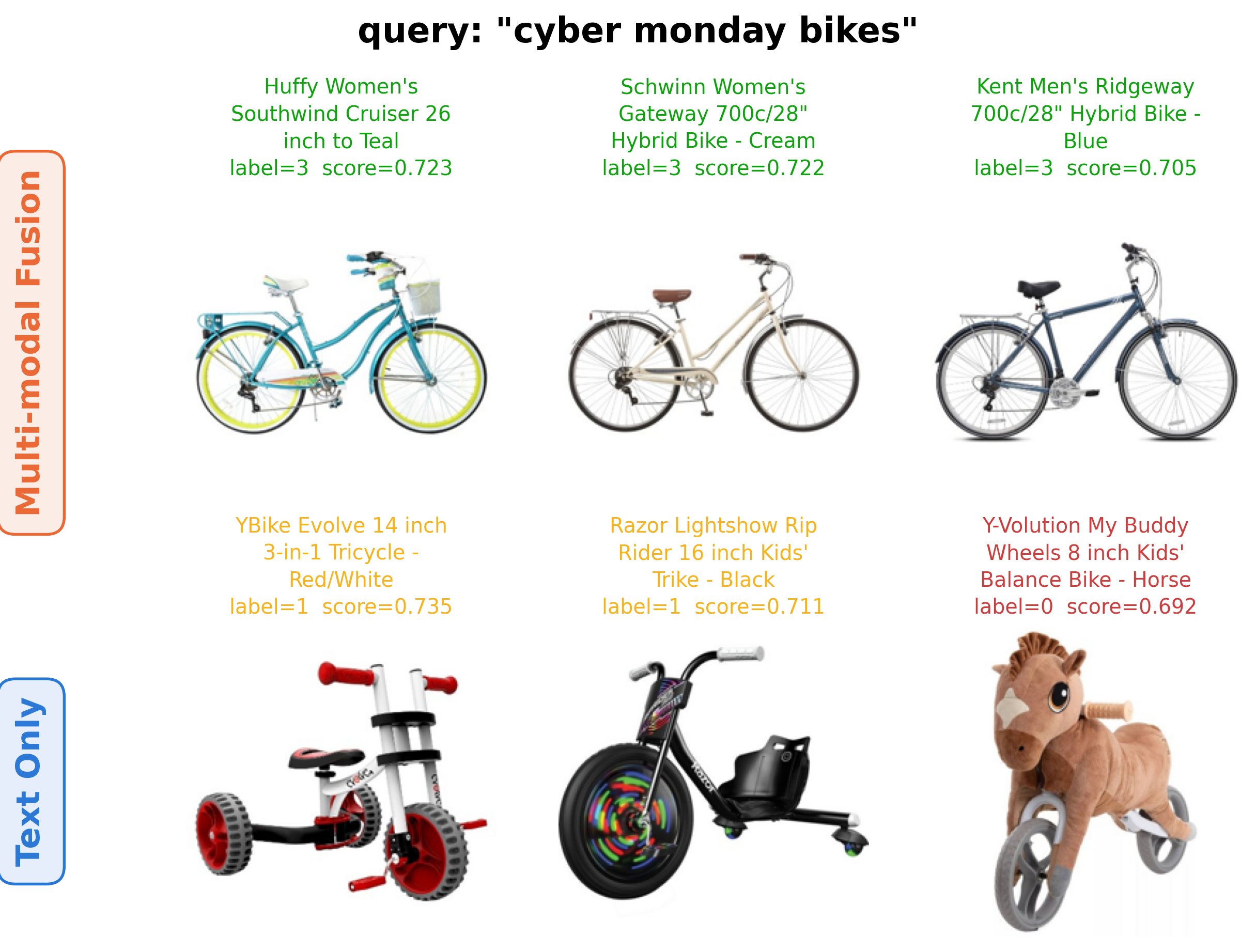}
\caption{Query \emph{``cyber monday bikes''} (product\_type). The multimodal model returns adult
bicycles (label~3); the text-only model returns kids' tricycles and a balance bike
(label~0--1)---a sub-type confusion that lexical overlap on ``bike'' cannot resolve.}
\label{fig:qual_bikes}
\end{figure}

\begin{figure}[t]
\centering
\includegraphics[width=0.95\columnwidth]{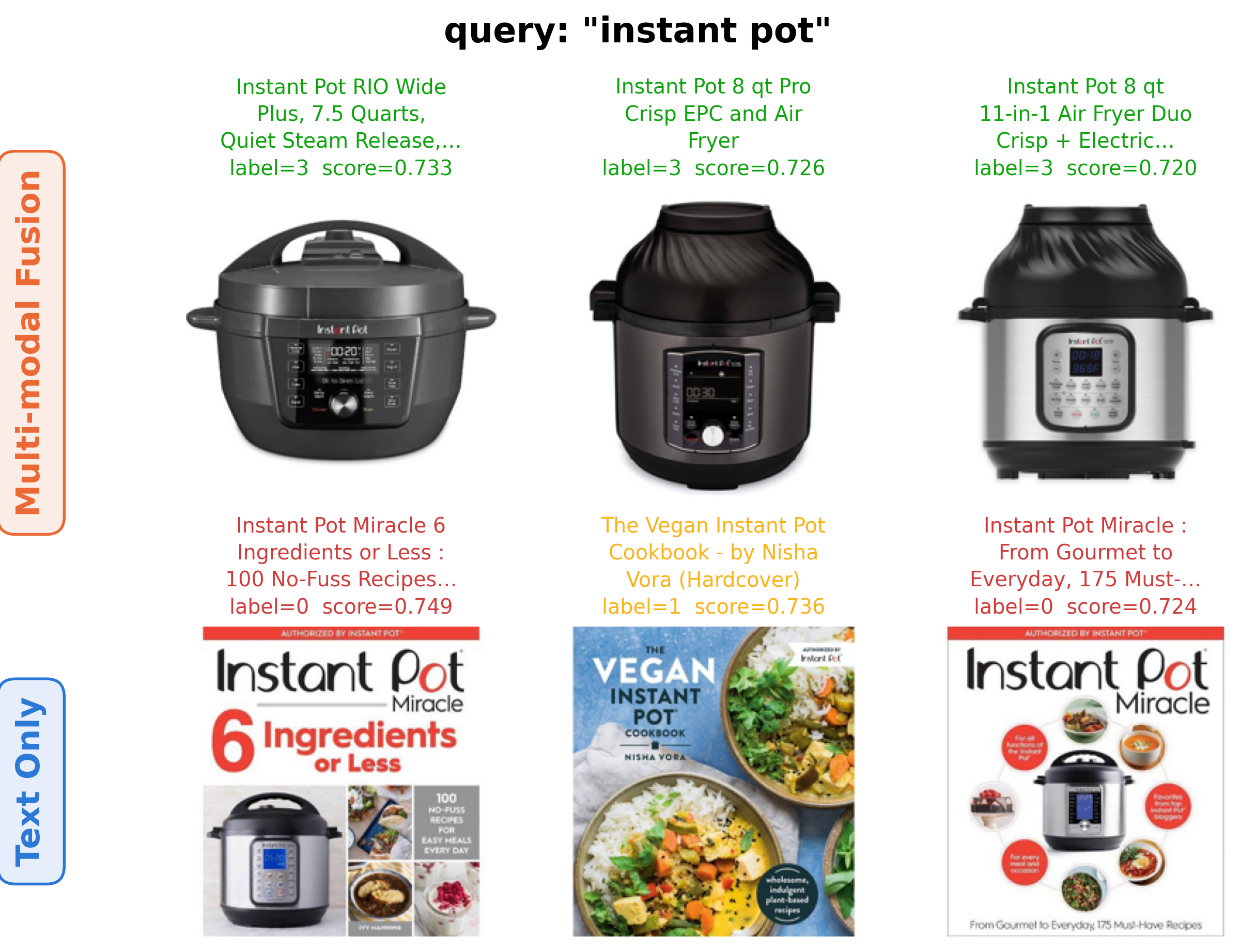}
\caption{Query \emph{``instant pot''} (branded). The multimodal model returns Instant Pot
appliances (label~3); the text-only model returns Instant Pot cookbooks (label~0--1)---visual
signal rescues even a branded query when a strong lexical match points to the wrong product
category.}
\label{fig:qual_instantpot}
\end{figure}

Figures~\ref{fig:qual_giftbag}, \ref{fig:qual_bikes}, and \ref{fig:qual_instantpot} illustrate this
pattern with concrete examples spanning three of the five intent buckets: a \emph{visual\_attribute}
query resolved by color (Fig.~\ref{fig:qual_giftbag}), a \emph{product\_type} query resolved by
sub-type appearance (Fig.~\ref{fig:qual_bikes}), and a \emph{branded} query where visual signal
overrides a misleading lexical match (Fig.~\ref{fig:qual_instantpot}). Together with the six
additional examples in the Appendix (Fig.~\ref{fig:qual_appendix_grid}), these cases show
that the gain from adding the image modality is not confined to a single query type, but recurs
across the intent categories where Table~\ref{tab:query_intent} shows the largest quantitative
improvements.

\subsection{Ablation Studies}
\label{sec:ablations}

\textbf{Ablation on Domain Fine-Tuning and Query Alignment.} We study the effect of different multimodal alignment
strategies.
Table~\ref{tab:alignment_ablation} compares three variants: pretrained CLIP, with domain
fine-tuning (+FT), and with additional query alignment (+FT+Align). Domain fine-tuning provides
modest but consistent improvements (e.g., +0.1\% Desirability@1), indicating better adaptation
to e-commerce semantics. Query alignment yields substantially larger gains. The +FT+Align variant improves
Desirability@1 by 1.7\% over CLIP and Relevance@1 by 0.2\%---highlighting the importance of
reducing the mismatch between user intent and item representations through explicit
modality-level alignment.

\textbf{Ablation on Fusion Networks.}
Table~\ref{tab:fusion_ablation} compares five multimodal fusion architectures. MoE+Bilinear
achieves the strongest overall performance on both desirability and relevance. Notably,
MoE+MLP matches MoE+Bilinear on Relevance (0.001 higher at @3) but falls short on Desirability (0.836 vs.\ 0.841
at @1): MLP fusion applies a symmetric transformation that cannot capture the asymmetric,
multiplicative dependencies between text and image features, whereas bilinear element-wise
multiplication naturally models such cross-modal interactions. These results confirm that both
adaptive modality weighting and expressive cross-modal interaction are necessary for optimal
multimodal retrieval.

\textbf{Ablation on Multi-Objective Loss Weights.}
We ablate the relevance/engagement weighting $\lambda_{\text{rel}}$ in \eqref{eq:total} using
the text-only two-tower model, since this axis is orthogonal to the fusion architecture ablated
above. Table~\ref{tab:weight_ablation} shows the deployed weighting ($\lambda_{\text{rel}}=0.3$)
matches or exceeds an even split ($\lambda_{\text{rel}}=0.5$) on \emph{both} evaluation
sets---not a compromise, but close to Pareto-best. Dropping relevance supervision entirely
($\lambda_{\text{rel}}=0$) trades a modest further gain on Desirability for a much larger loss on
Relevance, confirming that relevance supervision regularizes against over-fitting to engagement
signal (Section~\ref{sec:training_objective}).

\textbf{Analysis of Learned Fusion Behaviors.}
To better understand the learned multimodal behaviors, we analyze the fusion weights and
interaction outputs of the proposed MoE-based model across product categories. In categories
with visually similar products (e.g., apparel), the model assigns higher weight to textual
representations; in visually distinctive categories (e.g., consumer electronics), it relies more
on image representations. On average, text weight is 0.65, indicating the continued importance
of textual information. This is corroborated by our per-intent comparison
(Section~\ref{sec:visual_signal_analysis}): apparel is the only top-level category with a
non-positive mean $\Delta$nDCG@10 on Relevance ($\Delta=-0.0004$), confirming that visual
similarity is least discriminative precisely where the model learns to lean on text. Bilinear
interaction activations are stronger for products where relevance is jointly determined by text
and image cues (e.g., furniture with specific styles or functional attributes).

\begin{table}[t]
\centering
\footnotesize
\caption{Ablation on domain fine-tuning and query alignment (nDCG).
FT = domain fine-tuning on 4M title--image pairs;
Align = modality-specific query alignment (Stage~II).}
\label{tab:alignment_ablation}
\setlength{\tabcolsep}{3pt}
\begin{tabular}{lcccccccc}
\toprule
\multirow{2}{*}{\textbf{Variant}}
& \multicolumn{4}{c}{\textbf{Desirability}}
& \multicolumn{4}{c}{\textbf{Relevance}} \\
\cmidrule(lr){2-5} \cmidrule(lr){6-9}
& @1 & @3 & @9 & @24 & @1 & @3 & @9 & @24 \\
\midrule
CLIP
& 0.824 & 0.824 & 0.851 & 0.903
& 0.907 & 0.900 & 0.899 & 0.905 \\
+FT
& 0.825 & 0.828 & 0.853 & 0.904
& 0.908 & 0.902 & 0.899 & 0.906 \\
+FT+Align
& \textbf{0.838} & \textbf{0.838} & \textbf{0.860} & \textbf{0.908}
& \textbf{0.909} & \textbf{0.904} & \textbf{0.902} & \textbf{0.908} \\
\bottomrule
\end{tabular}
\end{table}

\begin{table}[t]
\centering
\footnotesize
\caption{Comparison of multimodal fusion architectures (nDCG).}
\label{tab:fusion_ablation}
\setlength{\tabcolsep}{3pt}
\begin{tabular}{lcccccccc}
\toprule
\multirow{2}{*}{\textbf{Fusion}}
& \multicolumn{4}{c}{\textbf{Desirability}}
& \multicolumn{4}{c}{\textbf{Relevance}} \\
\cmidrule(lr){2-5} \cmidrule(lr){6-9}
& @1 & @3 & @9 & @24 & @1 & @3 & @9 & @24 \\
\midrule
MLP          & 0.827 & 0.829 & 0.853 & 0.904 & 0.908 & 0.901 & 0.898 & 0.905 \\
MoE          & 0.838 & 0.838 & 0.860 & 0.908 & 0.909 & 0.904 & 0.902 & 0.908 \\
MoE+MLP      & 0.836 & 0.836 & 0.859 & 0.908 & 0.911 & 0.906 & 0.902 & 0.909 \\
Attention    & 0.833 & 0.834 & 0.857 & 0.906 & 0.908 & 0.903 & 0.900 & 0.907 \\
MoE+Bilinear & \textbf{0.841} & \textbf{0.839} & \textbf{0.861} & \textbf{0.909}
             & \textbf{0.911} & \textbf{0.905} & \textbf{0.903} & \textbf{0.909} \\
\bottomrule
\end{tabular}
\end{table}

\begin{table}[t]
\centering
\footnotesize
\caption{Ablation on multi-objective loss weights (nDCG), \textbf{text-only two-tower model}.
$\lambda_{\text{rel}}$ = weight on the relevance objective in \eqref{eq:total}
($\lambda_{\text{eng}}=1-\lambda_{\text{rel}}$).}
\label{tab:weight_ablation}
\setlength{\tabcolsep}{3pt}
\begin{tabular}{lcccccccc}
\toprule
\multirow{2}{*}{\textbf{Weighting}}
& \multicolumn{4}{c}{\textbf{Desirability}}
& \multicolumn{4}{c}{\textbf{Relevance}} \\
\cmidrule(lr){2-5} \cmidrule(lr){6-9}
& @1 & @3 & @9 & @24 & @1 & @3 & @9 & @24 \\
\midrule
$\lambda_{\text{rel}}=0.5$
& 0.779 & 0.787 & 0.823 & 0.885
& 0.889 & 0.888 & 0.890 & 0.897 \\
$\lambda_{\text{rel}}=0.3$
& \textbf{0.802} & \textbf{0.809} & \textbf{0.839} & \textbf{0.895}
& \textbf{0.890} & \textbf{0.888} & \textbf{0.890} & \textbf{0.897} \\
$\lambda_{\text{rel}}=0$
& 0.814 & 0.819 & 0.851 & 0.907
& 0.866 & 0.858 & 0.873 & 0.883 \\
\bottomrule
\end{tabular}
\end{table}

\subsection{Deployment Considerations}
\label{sec:deployment}

By fusing modalities entirely within the item tower, our architecture ensures that the image
encoder and bilinear interaction network are invoked only during offline embedding
computation---never at serving time. This design choice directly enables CPU-only production
deployment with zero additional \emph{online} serving overhead compared to the text-only
baseline. The added cost is offline: every product image must be encoded and fused when the
index is built or refreshed.

The system indexes approximately 8 million products, each represented by a 512-dimensional
fused embedding stored in a ScaNN~\cite{guo2020scann} ANN index. At query time, we run only the
lightweight CLIP text encoder, producing a query embedding dispatched to the index for top-$K$
retrieval; end-to-end latency stays under 100\,ms (p99) at roughly 2{,}000 queries per second on
commodity CPU-based infrastructure. The multimodal model thus deploys as a drop-in replacement
without changes to the existing serving stack.

We refresh item embeddings in batch mode: when updated products exceed a configurable threshold,
a batch job re-encodes affected items and atomically swaps the corresponding ANN index shards,
amortizing image-encoder cost across millions of items while keeping the online serving path
free of the heavier encoder. We monitor the deployed system with the same online metrics used
for the text-only baseline (click-through, conversion rate, and offline nDCG on held-out logs).

\section{Conclusion}

Our experiments demonstrate that bridging the multimodal gap in e-commerce retrieval---by
incorporating product images into a two-tower framework---yields consistent and substantial improvements in both user engagement and semantic relevance metrics. Domain-specific fine-tuning and explicit query alignment to text and image modalities are the key enablers, and our mixture-of-modality-experts architecture with bilinear interaction provides the most effective fusion strategy among the designs we evaluated. Critically, the fusion-at-item-tower design incurs zero additional online serving overhead: product embeddings are pre-computed offline, and the online serving path remains identical to the text-only baseline.

Beyond the aggregate metrics, our diagnostic analysis of the individual encoder towers
(Section~\ref{sec:visual_signal_analysis}) and our analysis of learned fusion behaviors
(Section~\ref{sec:ablations}) together paint a consistent picture: the model does not apply visual
signal uniformly, but learns to lean on the image tower precisely where text is least
discriminative---for color and material attributes, for sub-type distinctions within a broad
lexical match, and for categories where visually similar products carry different titles---while
continuing to rely on text where it remains the more reliable signal, such as apparel and brand
disambiguation. This selectivity, rather than a fixed image/text weighting, is what the
mixture-of-modality-experts gate and the bilinear interaction network are designed to capture, and
it is consistent with how a human shopper would weigh the two modalities differently depending on
the product category.

A limitation of this design is that, because images are fused only in the item tower, the query tower cannot yet accept an image as input, so purely visual queries (e.g., a photo of a desired product) are out of scope for the current system. Our diagnostic analysis also relies on lexical rules to bucket queries by intent (Section~\ref{sec:visual_signal_analysis}), which is a coarse proxy for true user intent and may under- or over-count some categories. These findings nonetheless provide practical guidance for building scalable multimodal retrieval systems in production, and point to two directions for future work: incorporating visual encoders into the query tower to support image-based queries, which requires addressing the latency constraints of online image encoding within the two-tower serving framework; and replacing the lexical query-intent taxonomy with a learned classifier to obtain a sharper picture of where visual signal helps most.

\section*{Acknowledgment}

We would like to thank Tianbao Yang and Xiyuan Wei from Texas A\&M University for their
insightful discussions and valuable guidance during this work. We also acknowledge their
open-source FastCLIP codebase, which provided helpful references and implementation insights
for our research (\url{https://github.com/Optimization-AI/FastCLIP}).

\suppressfloats[b]
\begin{figure*}[b]
{\centering\textsc{Appendix}\par}
\vspace{4pt}
\parbox{\textwidth}{\setlength{\parindent}{1em}\indent Section~\ref{sec:visual_signal_analysis} illustrates three representative queries where the
multimodal model outperforms the text-only model. Here we provide six additional examples
selected by the same procedure: for each query, we take the top-3 items ranked by the multimodal
model and by the text-only model independently, restricted to cases with no overlap between the
two top-3 sets and where the multimodal model's selection has a strictly higher mean label. Each
figure shows the multimodal model's top-3 (top row) and the text-only model's top-3 (bottom row),
with product title, ground-truth label, and similarity score annotated below each thumbnail.}
\vspace{8pt}

\centering
\begin{minipage}[t]{0.32\textwidth}
\centering
\includegraphics[width=\linewidth]{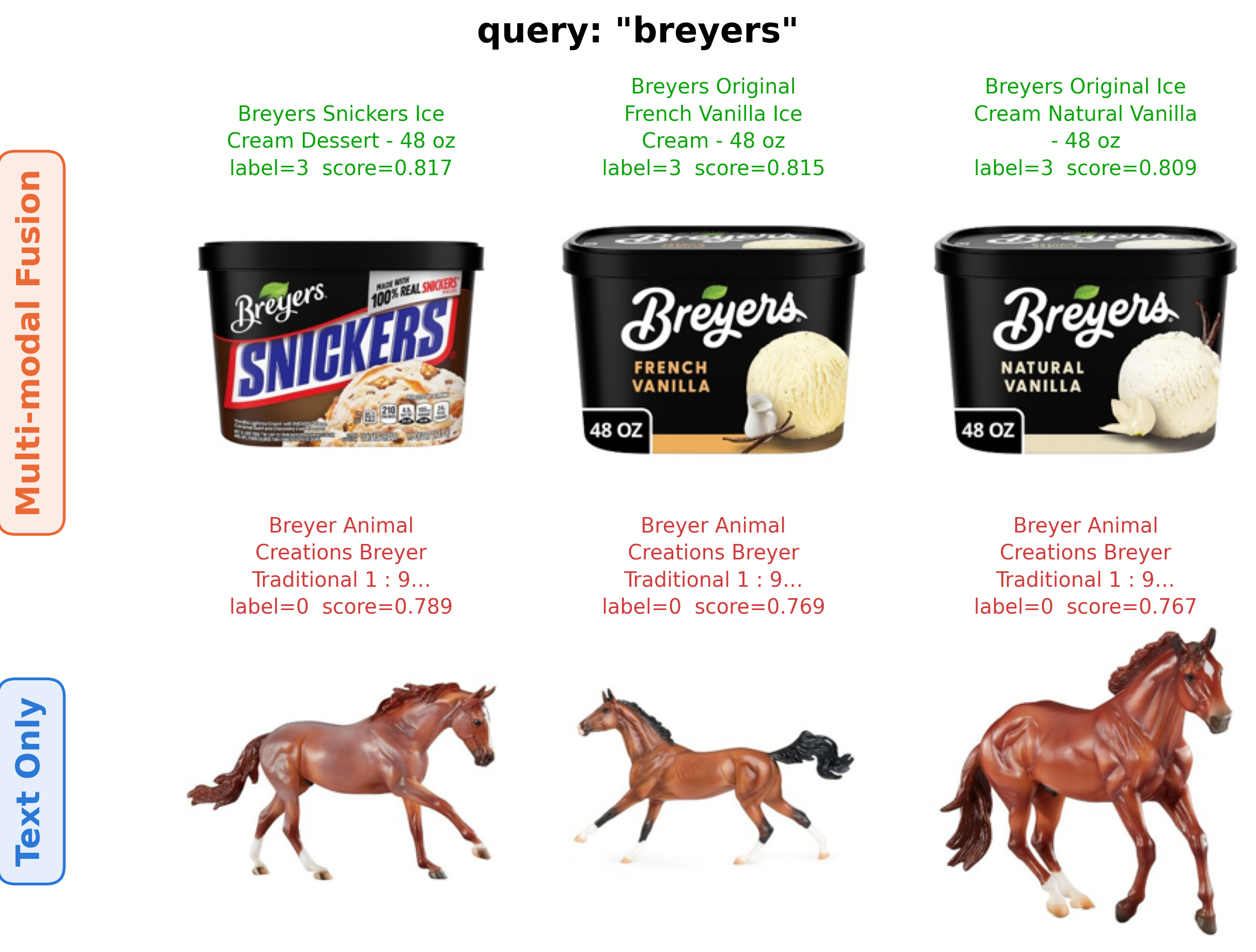}\\[2pt]
{\footnotesize\textbf{(a)} \emph{``breyers''} (generic\_broad): multimodal returns Breyers ice
cream (label~3); text-only returns unrelated Breyer toy figurines (label~0), matched purely on the
brand-name sub-string.}
\end{minipage}
\hfill
\begin{minipage}[t]{0.32\textwidth}
\centering
\includegraphics[width=\linewidth]{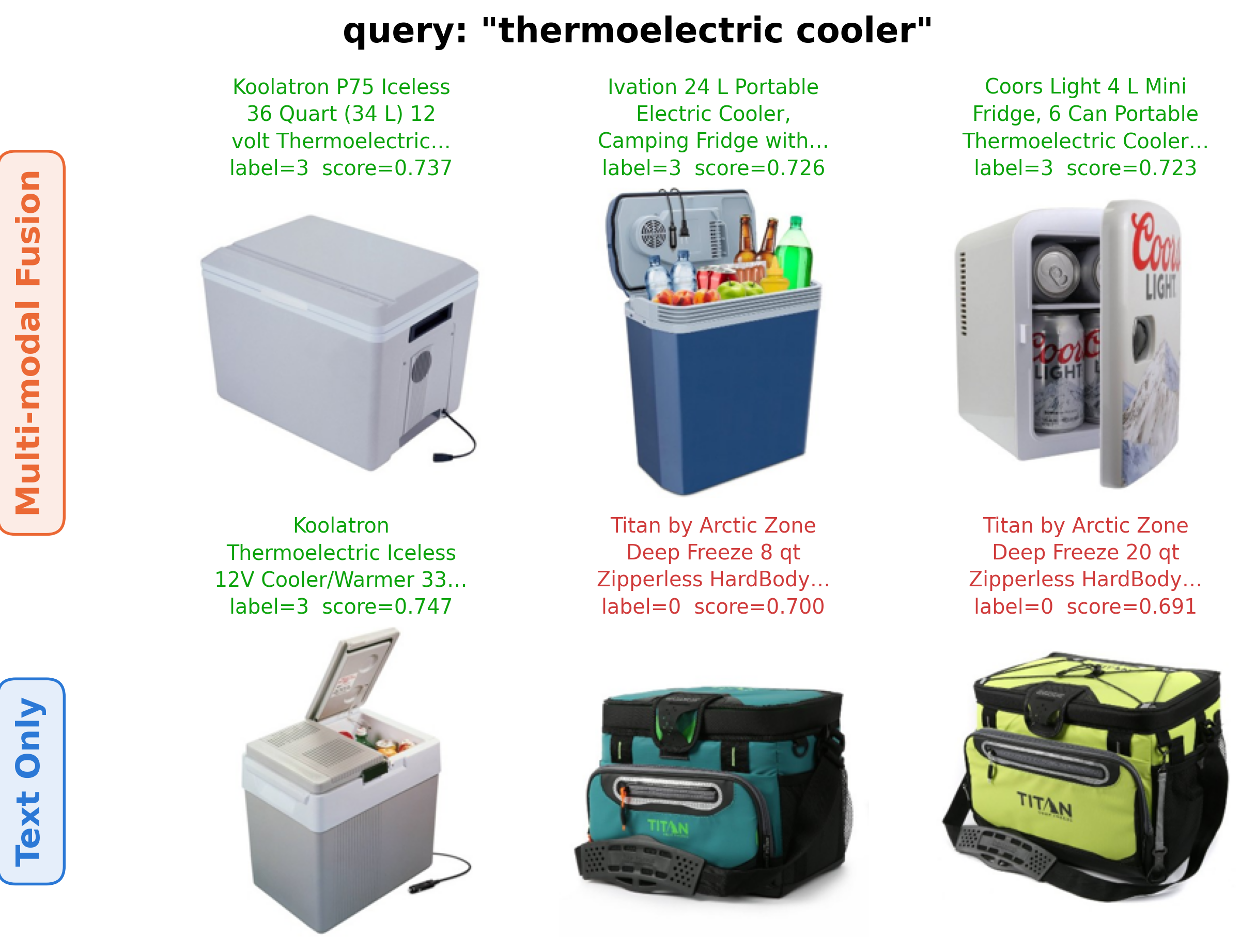}\\[2pt]
{\footnotesize\textbf{(b)} \emph{``thermoelectric cooler''} (generic\_broad): multimodal returns
actual thermoelectric coolers (label~3); text-only mixes in an unrelated zipperless cooler
(label~0) via lexical overlap on ``cooler.''}
\end{minipage}
\hfill
\begin{minipage}[t]{0.32\textwidth}
\centering
\includegraphics[width=\linewidth]{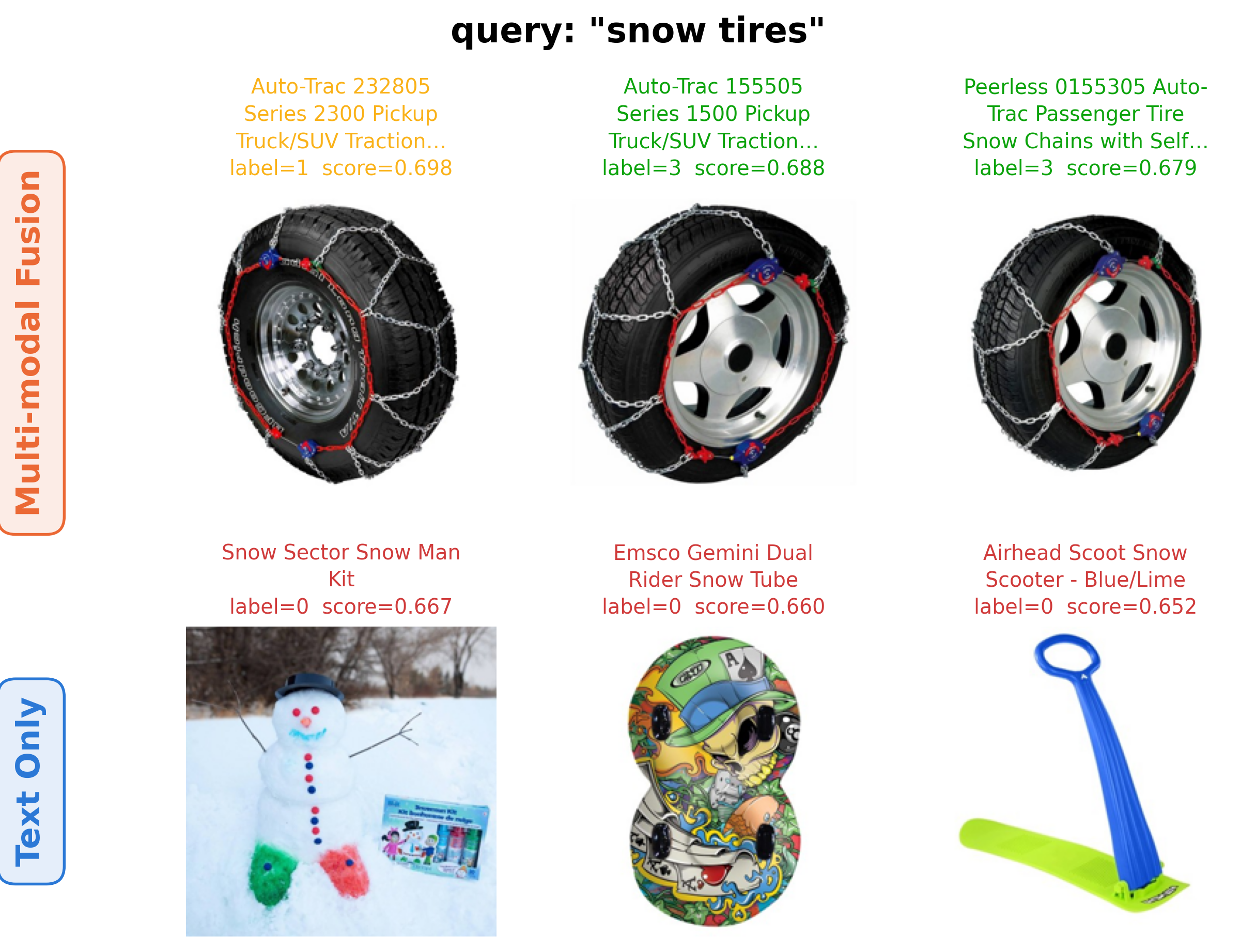}\\[2pt]
{\footnotesize\textbf{(c)} \emph{``snow tires''} (product\_type): multimodal returns tire chains
(label~1--3); text-only returns unrelated snow toys and scooters (label~0).}
\end{minipage}

\vspace{6pt}

\begin{minipage}[t]{0.32\textwidth}
\centering
\includegraphics[width=\linewidth]{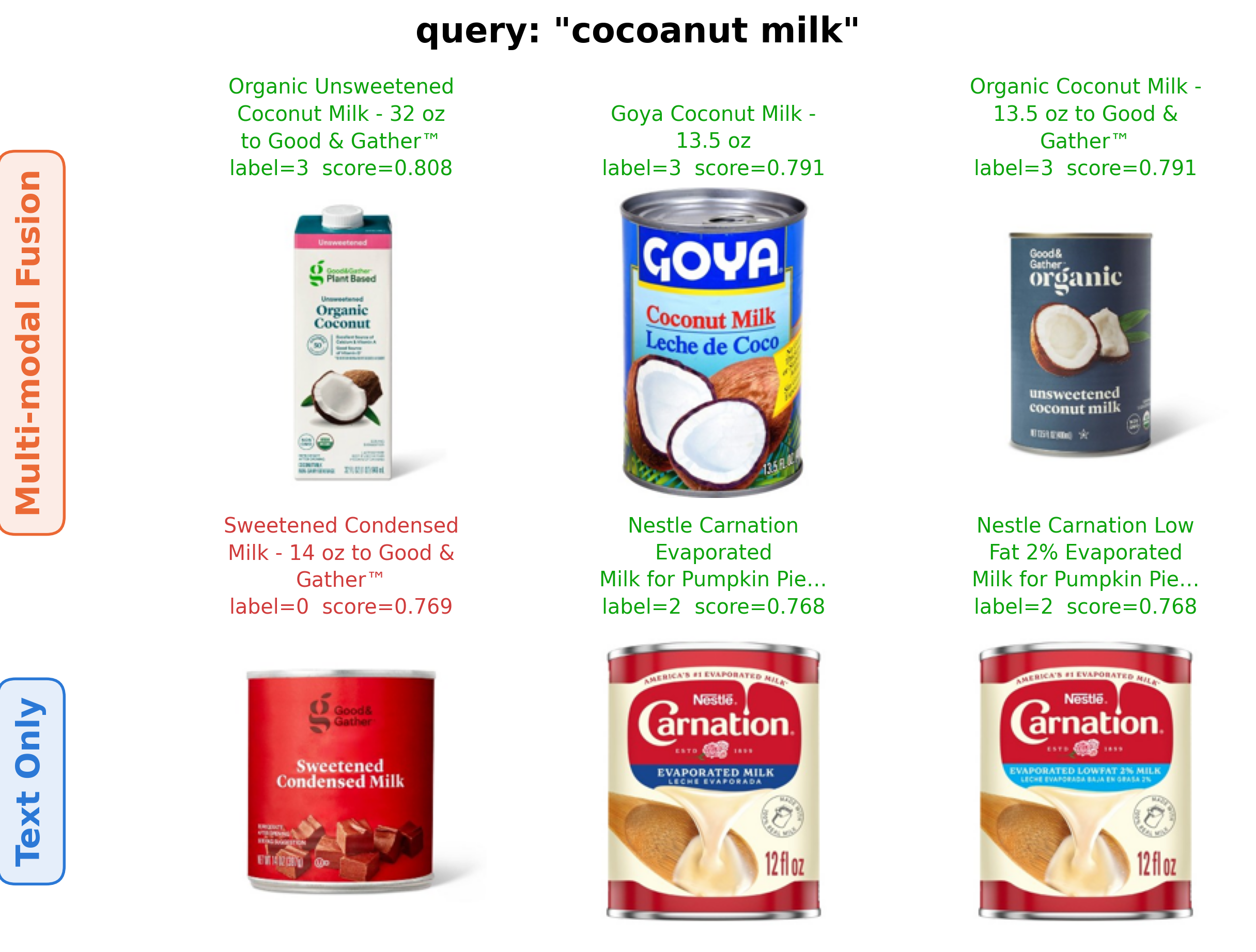}\\[2pt]
{\footnotesize\textbf{(d)} \emph{``cocoanut milk''} (product\_type): multimodal returns coconut
milk products (label~3); text-only returns unrelated condensed/evaporated milk (label~0--2),
matched on the shared word ``milk.''}
\end{minipage}
\hfill
\begin{minipage}[t]{0.32\textwidth}
\centering
\includegraphics[width=\linewidth]{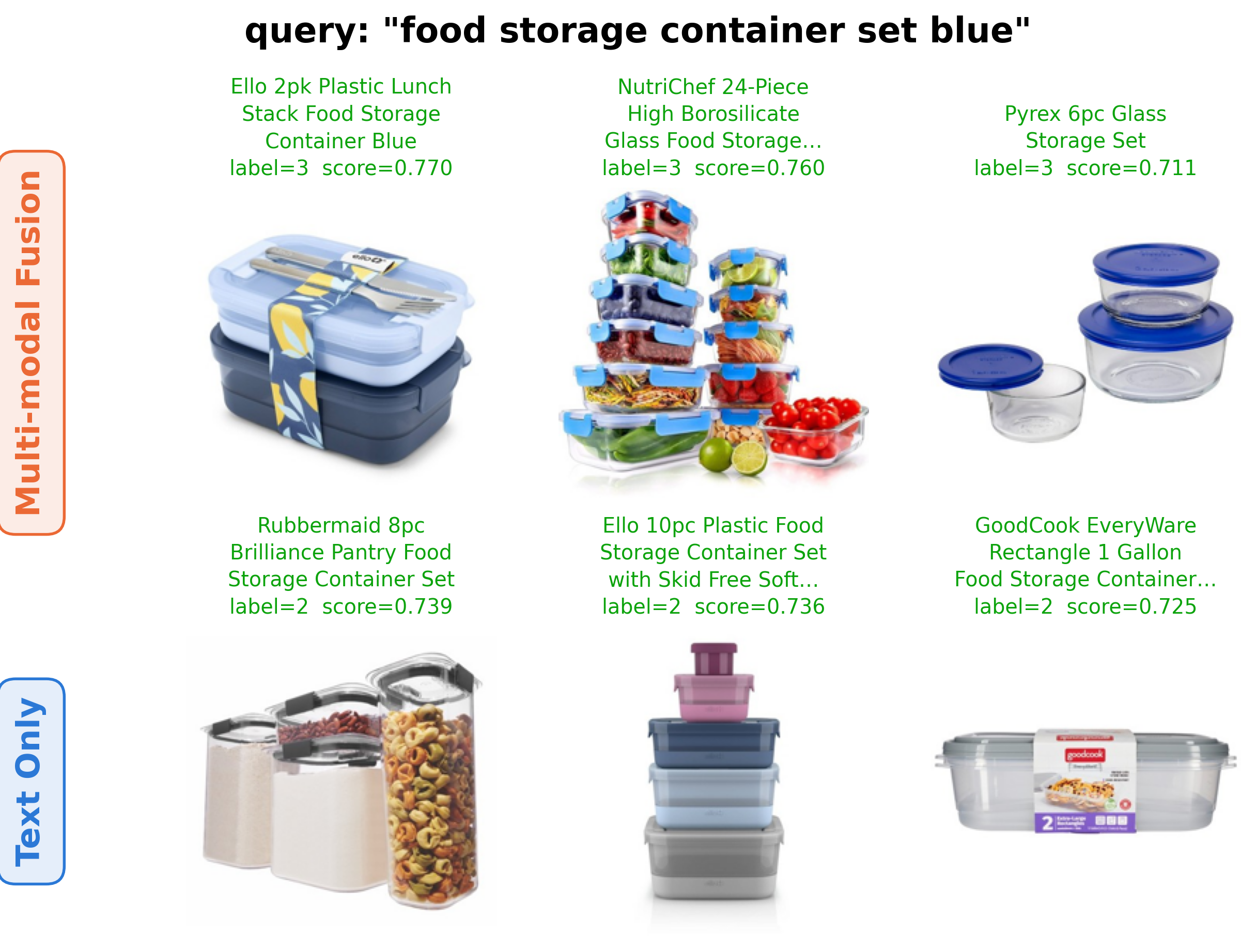}\\[2pt]
{\footnotesize\textbf{(e)} \emph{``food storage container set blue''} (visual\_attribute):
multimodal returns blue containers (label~3); text-only matches the correct product type but the
wrong color (label~2).}
\end{minipage}
\hfill
\begin{minipage}[t]{0.32\textwidth}
\centering
\includegraphics[width=\linewidth]{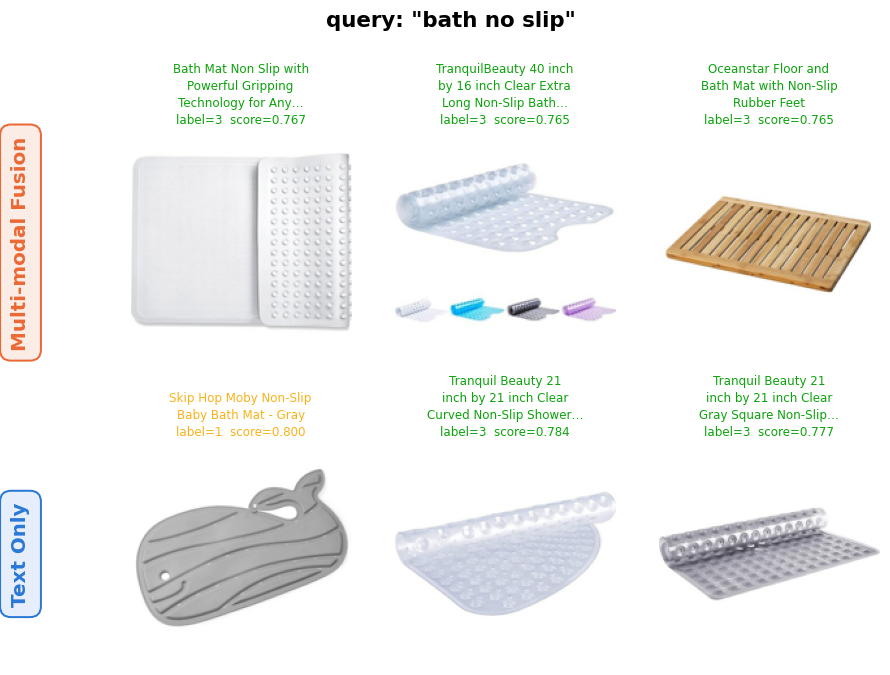}\\[2pt]
{\footnotesize\textbf{(f)} \emph{``bath no slip''} (functional\_benefit): multimodal returns
non-slip bath mats (label~3 for all three); text-only's top result is a lexically similar but
categorically different baby bath mat (label~1) ranked above two label-3 items.}
\end{minipage}
\caption{Six additional qualitative examples, selected by the same procedure as
Section~\ref{sec:visual_signal_analysis}: the multimodal model's top-3 (top row of each panel) vs.\
the text-only model's top-3 (bottom row of each panel), with product title, ground-truth label,
and similarity score annotated below each thumbnail.}
\label{fig:qual_appendix_grid}
\end{figure*}

\bibliographystyle{IEEEtran}
\bibliography{references}

@IEEEtranBSTCTL{IEEEexample:BSTcontrol,
  CTLdash_repeated_names = "no"
}

@inproceedings{Yu_KDD_2022_CommerceMM,
  title     = {{CommerceMM}: Large-scale commerce multimodal representation learning with omni retrieval},
  author    = {Yu, Licheng and Chen, Jun and Sinha, Animesh and Wang, Mengjiao and Chen, Yu and Berg, Tamara L and Zhang, Ning},
  booktitle = {Proceedings of the 28th ACM SIGKDD Conference on Knowledge Discovery and Data Mining},
  pages     = {4433--4442},
  year      = {2022}
}

@inproceedings{Jin_CIKM_2023_MSRA,
  title     = {MSRA: A multi-aspect semantic relevance approach for e-commerce via multimodal pre-training},
  author    = {Jin, Hanqi and Tan, Jiwei and Liu, Lixin and Qiu, Lisong and Yao, Shaowei and Chen, Xi and Zeng, Xiaoyi},
  booktitle = {Proceedings of the 32nd ACM International Conference on Information and Knowledge Management},
  pages     = {3988--3992},
  year      = {2023}
}

@inproceedings{radford2021clip,
  title   = {Learning Transferable Visual Models From Natural Language Supervision},
  author  = {Radford, Alec and Kim, Jong Wook and Hallacy, Chris and Ramesh, Aditya and Goh, Gabriel and Agarwal, Sandhini and Sastry, Girish and Askell, Amanda and Mishkin, Pamila and Clark, Jack and others},
  booktitle = {International Conference on Machine Learning (ICML)},
  year    = {2021},
  url     = {https://arxiv.org/abs/2103.00020}
}

@inproceedings{fukui2016mcb,
  title     = {Multimodal Compact Bilinear Pooling for Visual Question Answering and Visual Grounding},
  author    = {Fukui, Akira and Park, Dong Huk and Yang, Daylen and Rohrbach, Anna and Darrell, Trevor and Rohrbach, Marcus},
  booktitle = {Advances in Neural Information Processing Systems},
  year      = {2016},
  url       = {https://arxiv.org/abs/1606.01847}
}

@inproceedings{benyounes2017mutan,
  title     = {MUTAN: Multimodal Tucker Fusion for Visual Question Answering},
  author    = {Ben-Younes, Hedi and Cadene, R{\'e}mi and Thome, Nicolas and Cord, Matthieu},
  booktitle = {IEEE International Conference on Computer Vision},
  year      = {2017},
  url       = {https://arxiv.org/abs/1705.06676}
}

@inproceedings{bringing_multimodality_amazon,
  title     = {Bringing Multimodality to Amazon Visual Search System},
  author    = {Zhu, Xinliang and Huang, Sheng-Wei and Ding, Han and Yang, Jinyu and Chen, Kelvin and Zhou, Tao and Neiman, Tal and Xie, Ouye and Tran, Son and Yao, Benjamin and Gray, Douglas and Bindal, Anuj and Dhua, Arnab},
  booktitle = {Proceedings of the 30th ACM SIGKDD Conference on Knowledge Discovery and Data Mining (KDD '24)},
  year      = {2024},
  pages     = {6390--6399},
  publisher = {ACM},
  doi       = {10.1145/3637528.3671640}
}

@inproceedings{zhou2022simans,
  title     = {{SimANS:} Simple Ambiguous Negatives Sampling for Dense Text Retrieval},
  author    = {Kun Zhou and Yeyun Gong and Xiao Liu and Wayne Xin Zhao and Yelong Shen and Anlei Dong and Jingwen Lu and Rangan Majumder and Ji-Rong Wen and Nan Duan and Weizhu Chen},
  booktitle = {Proceedings of the 2022 Conference on Empirical Methods in Natural Language Processing: Industry Track},
  year      = {2022}
}

@inproceedings{lin2023prod,
  title     = {PROD: Progressive Distillation for Dense Retrieval},
  author    = {Zhenghao Lin and Yeyun Gong and Xiao Liu and Hang Zhang and Chen Lin and Anlei Dong and Jian Jiao and Jingwen Lu and Daxin Jiang and Rangan Majumder and Nan Duan},
  booktitle = {Proceedings of the ACM Web Conference 2023},
  year      = {2023}
}

@inproceedings{he2023capstone,
  title     = {{CAPSTONE:} Curriculum Sampling for Dense Retrieval with Document Expansion},
  author    = {Xingwei He and Yeyun Gong and A-Long Jin and Hang Zhang and Anlei Dong and Jian Jiao and Siu Ming Yiu and Nan Duan},
  booktitle = {Proceedings of the 2023 Conference on Empirical Methods in Natural Language Processing},
  year      = {2023}
}

@inproceedings{MultimodalSemanticRetrieval,
  title     = {Multimodal semantic retrieval for product search},
  author    = {Liu, Dong and Ramos, Esther Lopez},
  booktitle = {Companion Proceedings of the ACM Web Conference 2025},
  pages     = {2170--2175},
  year      = {2025}
}

@inproceedings{PictureWorthWords,
  title     = {One picture is worth a thousand words? The pricing power of images in e-Commerce},
  author    = {Naumzik, Christof and Feuerriegel, Stefan},
  booktitle = {Proceedings of The Web Conference 2020},
  pages     = {3119--3125},
  year      = {2020}
}

@inproceedings{Vl-clip,
  title     = {{VL-CLIP}: Enhancing multimodal recommendations via visual grounding and {LLM}-augmented {CLIP} embeddings},
  author    = {Giahi, Ramin and Yao, Kehui and Kollipara, Sriram and Zhao, Kai and Mirjalili, Vahid and Xu, Jianpeng and Biswas, Topojoy and Korpeoglu, Evren and Achan, Kannan},
  booktitle = {Proceedings of the Nineteenth ACM Conference on Recommender Systems},
  pages     = {482--491},
  year      = {2025}
}

@article{UniECS,
  title   = {UniECS: Unified Multimodal E-Commerce Search Framework with Gated Cross-modal Fusion},
  author  = {Liang, Zihan and Ma, Yufei and Qian, ZhiPeng and Dai, Huangyu and Wang, Zihan and Chen, Ben and Lei, Chenyi and Ding, Yuqing and Li, Han},
  journal = {arXiv preprint arXiv:2508.13843},
  year    = {2025}
}

@inproceedings{Ei-clip,
  title     = {{EI-CLIP}: Entity-aware interventional contrastive learning for e-commerce cross-modal retrieval},
  author    = {Ma, Haoyu and Zhao, Handong and Lin, Zhe and Kale, Ajinkya and Wang, Zhangyang and Yu, Tong and Gu, Jiuxiang and Choudhary, Sunav and Xie, Xiaohui},
  booktitle = {Proceedings of the IEEE/CVF Conference on Computer Vision and Pattern Recognition},
  pages     = {18051--18061},
  year      = {2022}
}

@inproceedings{Category2Image,
  title     = {Extending CLIP for Category-to-image Retrieval in E-commerce},
  author    = {Hendriksen, Mariya and Bleeker, Maurits and Vakulenko, Svitlana and Van Noord, Nanne and Kuiper, Ernst and De Rijke, Maarten},
  booktitle = {European Conference on Information Retrieval},
  pages     = {289--303},
  year      = {2022},
  organization = {Springer}
}

@inproceedings{Fashionklip,
  title     = {{FashionKLIP}: Enhancing e-commerce image-text retrieval with fashion multi-modal conceptual knowledge graph},
  author    = {Wang, Xiaodan and Wang, Chengyu and Li, Lei and Li, Zhixu and Chen, Ben and Jin, Linbo and Huang, Jun and Xiao, Yanghua and Gao, Ming},
  booktitle = {Proceedings of the 61st Annual Meeting of the Association for Computational Linguistics (Volume 5: Industry Track)},
  pages     = {149--158},
  year      = {2023}
}

@inproceedings{instcartebr,
  author    = {Yuqing Xie and Taesik Na and Xiao Xiao and Saurav Manchanda and Young Rao and Zhihong Xu and Guanghua Shu and Esther Vasiete and Tejaswi Tenneti and Haixun Wang},
  title     = {An Embedding-Based Grocery Search Model at Instacart},
  booktitle = {Proceedings of the ACM SIGIR Workshop on eCommerce (SIGIR eCom)},
  year      = {2022},
  publisher = {ACM},
  note      = {arXiv:2209.05555},
  url       = {https://sigir-ecom.github.io/ecom22Papers/paper_8392.pdf}
}

@misc{target_unified_ranking_26,
  title         = {Unified Learning-to-Rank for Multi-Channel Retrieval in Large-Scale E-Commerce Search},
  author        = {Gaydhani, Aditya and Xu, Guangyue and Kamath, Dhanush and Singh, Ankit and Li, Alex},
  year          = {2026},
  month         = feb,
  howpublished  = {arXiv preprint arXiv:2602.23530},
  eprint        = {2602.23530},
  archivePrefix = {arXiv},
  primaryClass  = {cs.IR},
  doi           = {10.48550/arXiv.2602.23530},
  url           = {https://arxiv.org/abs/2602.23530}
}

@inproceedings{walmart_semantic_retrieval_22_kdd,
  title     = {Semantic retrieval at Walmart},
  author    = {Magnani, Alessandro and Liu, Feng and Chaidaroon, Suthee and Yadav, Sachin and Reddy Suram, Praveen and Puthenputhussery, Ajit and Chen, Sijie and Xie, Min and Kashi, Anirudh and Lee, Tony and others},
  booktitle = {Proceedings of the 28th ACM SIGKDD Conference on Knowledge Discovery and Data Mining},
  pages     = {3495--3503},
  year      = {2022}
}

@inproceedings{guo2020scann,
  title={Accelerating Large-Scale Inference with Anisotropic Vector Quantization},
  author={Guo, Ruiqi and Sun, Philip and Lindgren, Erik and Geng, Quan and Simcha, David and Chern, Felix and Kumar, Sanjiv},
  booktitle={International Conference on Machine Learning (ICML)},
  year={2020}
}

@article{wei2024fastclip,
  title={{FastCLIP}: A Suite of Optimization Techniques to Accelerate {CLIP} Training with Limited Resources},
  author={Wei, Xiyuan and Ye, Fanjiang and Yonay, Ori and Chen, Xingyu and Sun, Baixi and Tao, Dingwen and Yang, Tianbao},
  journal={arXiv preprint arXiv:2407.01445},
  year={2024}
}

@inproceedings{xu2021zscl,
  title     = {Zero-Shot Compositional Concept Learning},
  author    = {Xu, Guangyue and Kordjamshidi, Parisa and Chai, Joyce},
  booktitle = {Proceedings of the 1st Workshop on Meta Learning and Its Applications to Natural Language Processing},
  pages     = {19--27},
  year      = {2021}
}

@article{xu2022prompting,
  title   = {Prompting Large Pre-trained Vision-Language Models for Compositional Concept Learning},
  author  = {Xu, Guangyue and Kordjamshidi, Parisa and Chai, Joyce},
  journal = {arXiv preprint arXiv:2211.05077},
  year    = {2022}
}

@inproceedings{xu2023metarevision,
  title     = {{MetaReVision}: Meta-Learning with Retrieval for Visually Grounded Compositional Concept Acquisition},
  author    = {Xu, Guangyue and Kordjamshidi, Parisa and Chai, Joyce},
  booktitle = {Findings of the Association for Computational Linguistics: EMNLP 2023},
  pages     = {12224--12236},
  year      = {2023}
}

@inproceedings{xu2024gipcol,
  title     = {{GIPCOL}: Graph-Injected Soft Prompting for Compositional Zero-Shot Learning},
  author    = {Xu, Guangyue and Chai, Joyce and Kordjamshidi, Parisa},
  booktitle = {Proceedings of the IEEE/CVF Winter Conference on Applications of Computer Vision (WACV)},
  pages     = {5774--5783},
  year      = {2024}
}

@phdthesis{xu2024thesis,
  title  = {Grounded Compositional Concept Learning},
  author = {Xu, Guangyue},
  school = {Michigan State University},
  year   = {2024}
}

\end{document}